\documentclass[aps,preprint]{revtex4}
\usepackage{amssymb,graphicx}

\begin{document}

\title{A quantum model for collective recoil lasing}

\author{R. Bonifacio, M.M. Cola and N. Piovella}

\affiliation{ Dipartimento di Fisica, Universit\`a degli Studi di
Milano, INFM and INFN, Via Celoria 16, Milano I-20133, Italy.}

\begin{abstract}
Free Electron Laser (FEL) and Collective Atomic Recoil Laser
(CARL) are described by the same model of classical equations for
properly defined scaled variables. These equations are extended to
the quantum domain describing the particle's motion by a
Schr\"{o}dinger equation coupled to a self-consistent radiation
field. The model depends on a single collective parameter $\bar
\rho$ which represents the maximum number of photons emitted per
particle. We demonstrate that the classical model is recovered in
the limit $\bar \rho\gg 1$, in which the Wigner function
associated to the Schr\"{o}dinger equation obeys to the classical
Vlasov equation. On the contrary, for $\bar \rho\le 1$, a new
quantum regime is obtained in which both FELs and CARLs behave as
a two-state system coupled to the self-consistent radiation field
and described by Maxwell-Bloch equations.
\end{abstract}

\maketitle

\section{Introduction}

Apparently very different systems as High-Gain Free Electron Laser
(FEL) \cite{BPN} and Collective Atomic Recoil Laser (CARL)
\cite{CARL} exhibit similar behaviors, showing self-bunching and
exponential enhancement of the emitted radiation. Originally
conceived in a semiclassical framework, they can be as well
described quantum-mechanically
\cite{PREP,BC,RB,Moore,Bigelow,CARLQ}. However, it is not
explicitly evident how to obtain the classical limit starting from
the quantum description. First attempts to give a quantum
description of the FEL have been proposed in the 80s, starting
from a canonical quantization of the $N$-particle Hamiltonian in
the Heisenberg picture, to study photon statistics and quantum
initialization from vacuum in the linear regime \cite{Becker,BC}.
In 1988, Preparata proposed a quantum field theory of FEL
\cite{PREP}, in which he has shown that, for $N \gg 1$, the FEL
dynamics is solved by a single-electron Schr\"{o}dinger equation
coupled to a self-consistent radiation mode. The same model has
been recently obtained to describe CARL from a Bose-Einstein
condensate (BEC) at zero temperature \cite{Moore,CARLQ}.
Furthermore, it has been also proved experimentally
\cite{MIT,kozuma,LENS} that CARL in a BEC exhibits quantum recoil
effects when the average recoil velocity remains less than the
photon recoil limit.

In this Letter we start from the classical model describing both
CARLs and FELs and we extend it to the quantum realm showing the
correspondence between the Preparata model \cite{PREP} and the
CARL-BEC model \cite{LENS}. In particular, it is possible to
derive an equation for the Wigner function of the $N$-particle
system. The Wigner function obeys to a finite difference equation
which reduces to the classical Vlasov equation \cite{BV} in the
limit in which the number of photons emitted per particle is much
larger than unity. In the opposite limit, both CARLs and FELs
behave as a two-state system \cite{MIT:PRL,Gatelli,Armenia}
described by the well-known Maxwell-Bloch equations \cite{MBE}.

\section{CARL-FEL model}
Apparently the physics of FEL and CARL appears to be quite
different. The first describes a relativistic high current
electron beam with energy $mc^2\gamma_0$, injected in a magnet
('wiggler') with a transverse, static magnetic field $B_w$ and
periodicity $\lambda_w$, which radiates in the forward direction
at the wavelength $\lambda\sim \lambda_w(1+a_w^2)/2\gamma_0^2$,
where $a_w=eB_w/mc^2k_w$ is the wiggler parameter and
$k_w=2\pi/\lambda_w$. Instead, CARL consists of a collection of
two level atoms in a high-Q ring cavity driven by a far-detuned
laser pump of frequency $\omega_p$ which radiates at the frequency
$\omega\sim\omega_p$ in the direction opposite to the pump. In
both cases the radiation process arises from a collective
instability which originates a symmetry breaking in the spatial
distribution, \textit{i.e.} a self-bunching of particles which
group in regions smaller than the wavelength.

It can be shown that, under suitable conditions and introducing
proper dimensionless variables, the dynamics of both FELs and
CARLs is described by the following Hamiltonian \cite{RB}
\begin{equation}
H=\sum_{j=1}^N \left[ \frac{p_j^2}{\bar \rho}
 +i\sqrt{\frac{\bar \rho}{2N}}\left(a^{\dag}e^{-i\theta_j}-{\rm h.c.}\right)
\right]-\frac{\delta}{\bar \rho}\; a^{\dag} a, \label{ham}
\end{equation}
where $\theta_j$ and $p_j$ are the phase operator of the $j$-th
particle and its conjugate momentum operator, obeying
$[\theta_j,p_{j'}]=i\delta_{jj'}$.  In Eq.(\ref{ham}), $a$ and
$a^{\dag}$ are annihilation and creation operators for the forward
radiation mode photon, with $[a,a^{\dag}]=1$. Notice that the
dynamics described by Eq.(\ref{ham}) depends only on the parameter
$\bar \rho$ and on the detuning $\delta$, properly defined for the
two systems:
\begin{itemize}
    \item For FELs, $\bar\rho=q\rho_{F}$ and $\delta=q(\gamma_0-\gamma_r)/\gamma_r$, where
$q=mc\gamma_r/\hbar k$,
$\gamma_r=\sqrt{(\lambda_w/2\lambda)(1+a_w^2)}$ is the resonant
energy and
$\rho_{F}=(1/\gamma_r)(a_w/4ck_w)^{2/3}(e^2n/m\epsilon_0)^{1/3}$
is the BPN parameter for a FEL \cite{BPN}, $\theta=(k_w+k)z-ck t$,
$p=q(\gamma-\gamma_0)/\gamma_r$ and $k=2\pi/\lambda$.
    \item For CARLs,
$\bar\rho=\rho_{C}$ and $\delta=(\omega_p-\omega)/\omega_R$, where
$\omega_R=2\hbar k^2/m$ is the recoil frequency,
$\rho_{C}=(S_0/\omega_R)^{2/3}(\omega
d^2n/2\hbar\epsilon_0)^{1/3}$,
$S_0=\Delta\Omega/[2(\Gamma^2+\Delta^2+\Omega^2)]$, $\Omega$ is
the pump Rabi frequency, $\Delta$ is the pump-atom detuning,
$\Gamma$ is natural decay constant of the atomic transition and
$d$ is the dipole matrix element \cite{BdS}. Finally, $\theta=2kz$
and $p=mv_z/2\hbar k$, where $v_z$ is the longitudinal atomic
velocity.
\end{itemize}
In these definitions, $n=N/V$ is the particle density in the
radiation volume $V$ and $m$ is the particle mass. Notice that in
both cases $\bar\rho$ scales as $n^{1/3}$, i.e. as the reciprocal
of the inter-particle distance. Introducing $\bar p_j=(2/\bar
\rho)p_j$ and $A=(2/N\bar \rho)^{1/2}a$, the Heisenberg equations
associated with Eq.(\ref{ham}) are \cite{BPN,CARL}:
\begin{eqnarray}
\frac{d\theta_j}{d\tau}&=&\bar{p}_j \label{C1}\\
\frac{d\bar{p}_j}{d\tau}&=&-\left(Ae^{i\theta_j}+\rm{c.c.}\right)\label{C2}\\
\frac{dA}{d\tau}&=&\frac{1}{N}\sum_{i=1}^N
e^{-i\theta_j}+\frac{i\delta}{\bar\rho}A \label{C3},
\end{eqnarray}
where $j=1\ldots N$ and $\tau=\omega_R\rho_C t$ for CARL whereas
$\tau=(4\pi\rho_F/\lambda_w)z$ for FEL. Considering the operators
$\theta_j$, $\bar p_j$ and $A$ in Eqs.(\ref{C1})-(\ref{C3}) as
c-numbers, one obtain the well-known classical description for FEL
and CARL. The scaling of Eqs.(\ref{C1})-(\ref{C3}) is called
`universal' in the sense that, assuming resonance (\textit{i.e.}
$\delta=0$), the equations do not contain any parameter. Hence,
the scaling law of the various physical quantities can be obtained
from their definition in term of $\bar \rho$. In particular, from
the definition of $A$ and $\bar p$ it follows that the photon
number per particle and the momentum recoil are proportional to
$\bar \rho$ , both for FELs and CARLs. Hence, it is expected that
when $\bar \rho\gg 1$ the system behaves classically, whereas for
$\bar \rho\le 1$ quantum effects becomes relevant. Notice that
$|A|^2+N^{-1}\sum_j\bar p_j$ is a constant of motion in
Eqs.(\ref{C1})-(\ref{C3}), \textit{i.e.} the radiated intensity is
due to the average recoil.

\section{Quantum CARL-FEL model}
In ref. \cite {PREP} Preparata, using quantum field theory, has
shown that the collective dynamics of the system of $N\gg 1$
electrons in an FEL can be described by means of a single complex
scalar quantum field whose behavior is governed by a
Schr\"{o}dinger-type equation in the self-consistent radiation
field, which originates a pendulum-like potential:
\begin{eqnarray}
i\frac{\partial\psi}{\partial\tau}&=&-\frac{1}{\bar \rho}
\frac{\partial^2\psi}{\partial\theta^2}-\frac{i\bar \rho}{2}
\left[A e^{i\theta}- {\rm c.c.}\right]\psi
\label{psi}\\
\frac{dA}{d\tau}&=&\int_{0}^{2\pi}
d\theta|\psi|^2e^{-i\theta}+\frac{i\delta}{\bar\rho}A, \label{a}
\end{eqnarray}
where $\psi$ is normalized to one, i.e. $\int_{0}^{2\pi}
d\theta|\psi(\theta,\tau)|^2=1$. Note that Eq.(\ref{psi}) is the
Schr\"{o}dinger equation associated to the Hamiltonian (\ref{ham})
and Eq.(\ref{a}) corresponds to Eq.(\ref{C3}) when the classical
average of $e^{-i\theta}$ is replaced by the quantum ensemble
average. Quoting ref. \cite{PREP}, Eqs.(\ref{psi}) and (\ref{a})
are derived if one ``formulate the many-electron problem in the
language of quantum field theory and uses the large number $N$ of
electrons to evaluate the resulting path integral by saddle-point
techniques''. Recently, the same model of Eqs.(\ref{psi}) and
(\ref{a}) has been used to describe CARL from a BEC
\textbf{\cite{Moore,CARLQ,LENS}}. Hence, we propose the nonlinear
system of Eqs.(\ref{psi}) and (\ref{a}) as the quantum extension
of the CARL-FEL classical model. We now show that the classical
equations (\ref{C1})-(\ref{C3}) are recovered in the limit
$\bar\rho\gg 1$.

\section{Wigner function approach}

Let's consider the standard definition of the Wigner function for
a state with wave function $\psi(\theta,\tau)$:
\begin{equation}
W(\theta,p,\tau) =\frac{1}{2\pi}\int_{-\infty}^{+\infty}\!d\xi \,
e^{i\xi p}\, \psi^{*}\left(\theta-\frac{\xi}{2} ,\tau\right)
\psi\left(\theta+\frac{\xi}{2} ,\tau\right), \label{defW}
\end{equation}
so that
\begin{equation}\label{w:psi}
\int_{-\infty}^{+\infty} dp \;
W(\theta,p,\tau)=|\psi(\theta,\tau)|^2.
\end{equation}
One can show that Eq.(\ref{psi}) is equivalent to the following
finite difference equation for the quasi-probability distribution
$W(\theta,\bar p,\tau)$:
\begin{equation}
\label{Weqn} \frac{\partial W(\theta,\bar p,\tau)}{\partial \tau}+
\bar p \frac{\partial W(\theta,\bar p,\tau)}{\partial \theta} -
\frac{\bar \rho}{2} \left[ A e^{i\theta} +  {\rm c.c.}\right]
\left[ W\left(\theta,\bar p+\frac{1}{\bar \rho}, \tau\right) -
W\left(\theta,\bar p - \frac{1}{\bar \rho}, \tau\right) \right] =
0.
\end{equation}
Using Eq. (\ref{w:psi}), Eq. (\ref{a}) becomes:
\begin{equation}\label{a:W}
    \frac{dA}{d\tau}=\int_{-\infty}^{+\infty}d\bar p
\int_{0}^{2\pi}d\theta W(\theta,\bar p,\tau)e^{-i\theta}+
\frac{i\delta}{\bar \rho} A .
\end{equation}
We underline again that Eqs.(\ref{Weqn}) and (\ref{a:W}) are
equivalent to Eqs.(\ref{psi}) and (\ref{a}) using the Wigner
function representation. In the right hand side of
Eq.~(\ref{Weqn}), the incremental ratio $[W(\theta,\bar
p+\epsilon)-W(\theta,\bar p-\epsilon)]/(2\epsilon)\rightarrow
\partial W(\theta,\bar p)/\partial\bar p$ when
$\epsilon=1/\bar \rho\rightarrow 0$. Hence, for $\bar \rho\gg 1$
Eq.~(\ref{Weqn}) becomes the Vlasov equation:
\begin{equation}
\label{Vlasov} \frac{\partial W(\theta,\bar p,\tau)}{\partial
\tau} + \bar p \frac{\partial W(\theta,\bar p,\tau)}{\partial
\theta} - \left[ A e^{i\theta} + {\rm c.c.}\right] \frac{\partial
W(\theta,\bar p,\tau)}{\partial\bar p} = 0.
\end{equation}
Eqs. (\ref{a:W}) and (\ref{Vlasov}) are equivalent to the
classical Eqs. (\ref{C1})-(\ref{C3}). This means that the
particles behave classically, following a Newtonian motion, when
$\bar \rho\gg 1$ i.e. when the average number of photons scattered
per particle is much larger than unity. In this limit, the quantum
recoil effects due to the single photon scattering process is
negligible. On the contrary, a quantum regime of CARL or FEL
occurs when $\bar \rho\le 1$, in which each particle scatters only
one photon.
\begin{figure}
\begin{center}
\includegraphics[width=0.9\textwidth]{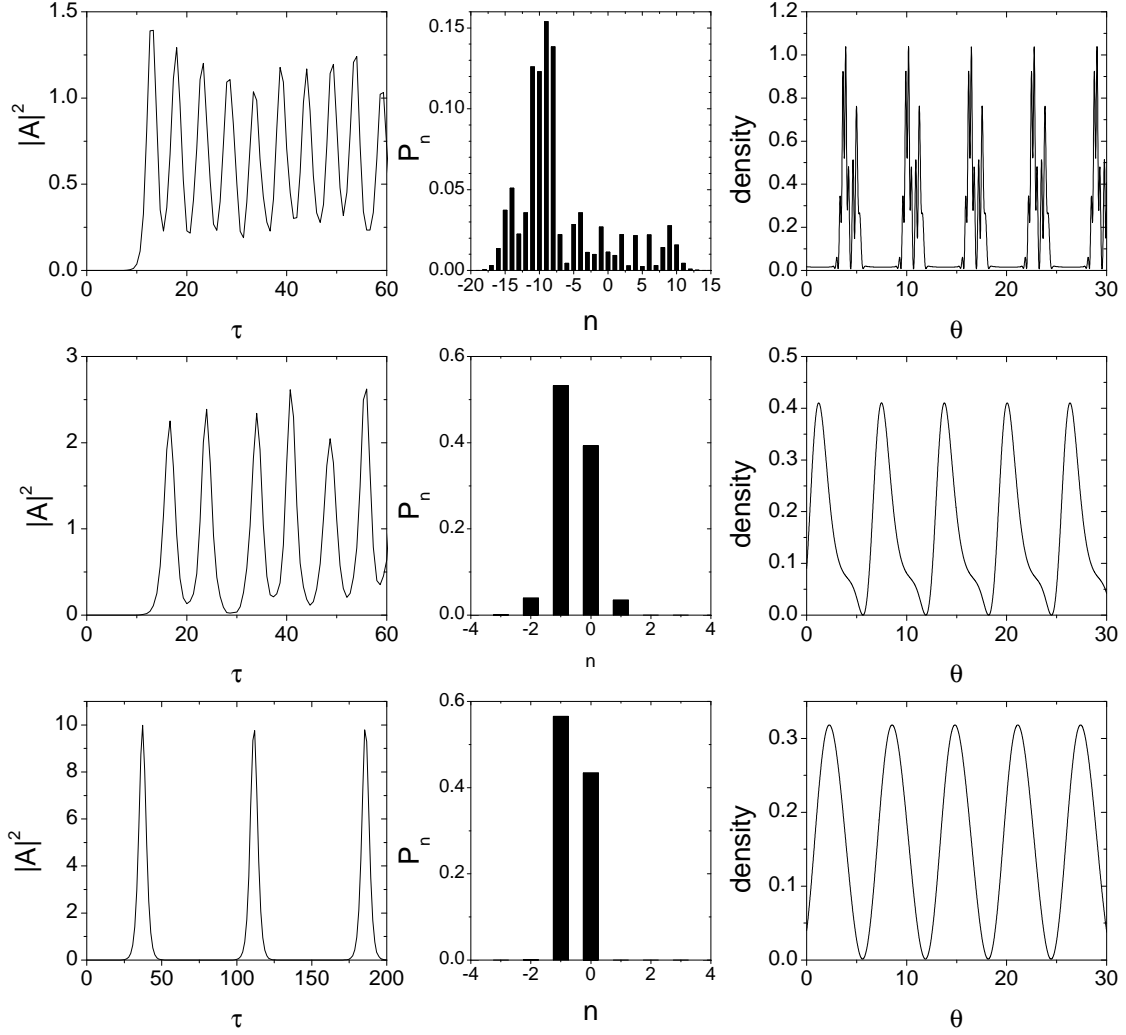}
\caption{Numerical
solution of Eq.(\ref{W3}) and (\ref{a3}) for $\bar \rho=10$ (first
row), $\bar \rho=1$ (second row) and $\bar \rho=0.2$ (third row).
The other parameters are $\delta=1$, $A(0)=10^{-4}$ and
$c_n(0)=\delta_{n0}$. Left column: dimensionless radiation
intensity $|A|^2$ vs. $\tau$; central and right column: occupation
probabilities $P_n=|c_n|^2$ vs. $n$ and density distribution
$|\psi|^2$ vs. $\theta$, for $\tau$ near the first maximum of
$|A|^2$.} \label{figura1} \end{center}
\end{figure}
In fact, expanding the wave function in Fourier series as
\begin{equation}
\psi(\theta,\tau)=\frac{1}{\sqrt{2\pi}}\sum_n
c_n(\tau)e^{in(\theta+\frac{\delta}{\bar \rho}\tau)}, \qquad
n=-\infty,\ldots , +\infty.
\end{equation}
and inserting this ansatz in Eqs.(\ref{psi}) and (\ref{a}), one
can easily obtain the following closed set of equations for
$\varrho_{m,n}(\tau)=c_{m}(\tau)^{*}c_n(\tau)$:
\begin{eqnarray}
\frac{d\varrho_{m,n}}{d\tau}&=& \frac{i}{\bar
\rho}(m-n)\left(\delta+m+n\right)\varrho_{m,n}+\frac{\bar \rho}{2}
\left[\bar{A}\left(\varrho_{m+1,n}-\varrho_{m,n-1}\right)+
\bar{A}^*\left(\varrho_{m,n+1}-\varrho_{m-1,n}\right)\right]\label{W3}\\
\frac{d\bar{A}}{d\tau}&=&\sum_{n=-\infty}^{\infty}\varrho_{n-1,n},
\label{a3}
\end{eqnarray}
where $\bar A= A e^{-i(\delta/\bar \rho)\tau}$. These equations
are equivalent to Eqs.(\ref{psi}) and (\ref{a}) for the density
matrix in the momentum representation and have been discussed in
ref.\cite{Gatelli}. Fig.\ref{figura1} shows the numerical solution
of Eq.(\ref{W3}) and (\ref{a3}) for $\bar \rho=10$ (first row),
$\bar \rho=1$ (second row) and $\bar \rho=0.2$ (third row). The
other parameters are $A(0)=10^{-4}$, $c_n(0)=\delta_{n0}$ and
$\delta=1$, which corresponds to a single photon scattering
recoil. In the first column $|A|^2$ is plotted as a function of
$\tau$. The central column shows the occupation probabilities
$P_n=|c_n|^2$ vs. $n$, whereas the right column shows the density
distribution $|\psi|^2$ vs. $\theta$, for a value of $\tau$ near
the first maximum of $|A|^2$. We note that for $\bar \rho=10$ the
system behaves classically: the momentum states are occupied in a
range of the order of $\bar \rho$ and the average momentum is
$\langle p\rangle\approx-\bar \rho$, as can be seen from the first
row of fig.\ref{figura1}. Furthermore, the particle distribution
shows periodic narrow peaks of density. When $\bar \rho=1$, the
mainly occupied momentum states are those for $n=0$ and $n=-1$,
corresponding to particles in the initial state or in the recoil
state, respectively. Finally, when $\bar \rho\ll 1$, the dynamics
is that of a pure two-level system. In this limit, if the
initially occupied state is the $n$th momentum state, the only two
momentum states involved in the interaction are those for $n$ and
$n-1$, so that Eq.(\ref{W3}) and (\ref{a3}), after defining the
`polarization' $S_n=2\varrho_{n-1,n}$ and the `population
difference' $D_n=\varrho_{n,n}-\varrho_{n-1,n-1}$, reduce to the
Maxwell-Bloch equations for a two-state system \cite{MBE}:
\begin{eqnarray}
\frac{dS_n}{d\tau'}&=& -i\Delta_n S_n
+A'D_n\label{S}\\
\frac{dD_n}{d\tau'}&=& -\frac{1}{2}\left(A'S_n^* + A^{'*}S_n\right)\label{W}\\
\frac{dA'}{d\tau'} &=& S_n \label{a:4}.
\end{eqnarray}
where $\Delta_n=(\delta-1+2n)/\bar \rho^{3/2}$, $A'=\sqrt{\bar
\rho}\bar{A}$ and $\tau'=\sqrt{\bar \rho}\tau$. With this new
scaling and assuming resonance (i.e. $\Delta_n=0$),
Eqs.(\ref{S})-(\ref{a:4}) do not contain any parameter. Hence, the
characteristic timescale is ruled by $\sqrt{n}$ instead of
$n^{1/3}$ as in the classical case. The quantum regime for CARLs
and FELs is analog to the coherent spontaneous emission regime
predicted quantum-mechanically for a two-level system in ref.
\cite{BP}, where a series of optical ``$2\pi$-pulses'' are
generated. In fact, assuming resonance (i.e. $\Delta_n=0$), $A'$
and $S_n$ are real. Hence, we can introduce the ``Bloch angle''
$\phi$ such that $S_n=\sin\phi$, $D_n=\cos\phi$. Then,
Eqs.(\ref{S})-(\ref{a:4}) reduce to a pendulum equation
$d^2\phi/d\tau'^2=\sin\phi$ and $d\phi/d\tau'=A'$. Hence, in the
quantum regime, the dynamics is that of a pendulum moving away
from the unstable equilibrium point ($\phi=0$) and undergoing
periodically a complete revolution ('$2\pi$-pulse') with angular
velocity $A'$.

Finally, we note that, adopting the same scaling of
Eqs.(\ref{S})-(\ref{a:4}) in Eq.(\ref{psi}), this can be
interpreted as a Schr\"{o}dinger equation for a single particle
with a ``mass'' $\bar \rho^{3/2}$ in a self-consistent pendulum
potential. This provides an intuitive interpretation of the
classical limit, that holds when the particle's 'mass' is large.
The strong differences between the quantum and classical regimes
are evident from Fig.\ref{figura1}.

\section{Conclusions}

In this Letter we presented a unified quantum model that stands
for apparently very different systems as FEL and CARL. The
dynamics is described by a Schr\"{o}dinger equation in a
self-consistent pendulum potential and is ruled by an unique
parameter $\bar \rho$ which represents the maximum number of
photons scattered per particle and the maximum momentum recoil in
units of the photon recoil momentum. The Schr\"{o}dinger equation
can be transformed in an exact equation for the Wigner
quasi-probability distribution. The main results are the
following: i) The classical model is recovered in the limit $\bar
\rho \gg 1$; this because the finite difference equation for the
Wigner function reduces to the classical Vlasov equation. ii) In
the limit $\bar \rho \le 1$ a completely different dynamical
regime occurs (see Fig. \ref{figura1}): due to momentum
quantization the system reduces to only two momentum states
obeying to the Maxwell-Bloch equations which describe the dynamics
of a two-level atomic system coupled to a coherent field.

\section{Acknowledgements}

One of us (RB) has to acknowledge a big mistake he made almost 20
years ago when Giuliano Preparata, a well-known field theorist,
presented to him his general ``Quantum Field Theory of a Free
Electron Laser'' \cite{PREP}. RB did not understand this work
thinking it was incorrect. On the contrary, Preparata's theory was
perfectly correct, as recognized  in this Letter, which is
dedicated to his memory.


\begin{thebibliography}{99}
\bibitem{BPN}  BONIFACIO R., PELLEGRINI C. and NARDUCCI L., Opt. Commun. 50 (1984) 373.
\bibitem{CARL} BONIFACIO R., DE SALVO SOUZA L., Nucl. Instrum. and Meth. in Phys.
Res. A 341 (1994) 360; BONIFACIO R., DE SALVO SOUZA L., NARDUCCI
L. and D'ANGELO E.J., Phys. Rev.A 50 (1994) 1716.
\bibitem{PREP} PREPARATA G., Phys. Rev.A 38 (1988) 233.
\bibitem{BC} BONIFACIO R. and CASAGRANDE F., Optics Comm. 50 (1984) 251.
\bibitem{RB} BONIFACIO R., Optics Comm. 146 (1998) 236.
\bibitem{Moore} MOORE M.G., ZOBAY O. and MEYSTRE P., Phys. Rev.A 60 (1999) 1491.
\bibitem{Bigelow} LING H.Y., PU H., BAKSMATY L. and BIGELOW N.P., Phys. Rev.A 63 (2001) 053810.
\bibitem{CARLQ} PIOVELLA N., COLA M., BONIFACIO R., Phys. Rev.A 67 (2003) 013817.
\bibitem{Becker} BECKER W. and MCIVER J.K., Phys. Rev.A 28 (1983) 1838.
\bibitem{LENS} BONIFACIO R., CATALIOTTI F.S., COLA M., FALANI L., FORT C., PIOVELLA N., INGUSCIO M., Optics Comm. 233 (2004) 155.
\bibitem{MIT} INOUYE S. et al., Science 285 (1999) 571.
\bibitem{kozuma} KOZUMA M. et al., Science 286 (1999) 2309; INOUYE S. et al., Nature
402 (1999) 641.
\bibitem{BV} BONIFACIO R., VERKERK P., Optics Comm. 124 (1996) 489.
\bibitem{BdS} BONIFACIO R. and DE SALVO SOUZA L., Optics Comm. 115 (1995) 505.
\bibitem{MIT:PRL} INOUYE S. et al., Phys. Rev. Lett. 85 (2000) 4225.
\bibitem{Gatelli} PIOVELLA N., GATELLI M. and BONIFACIO R., Optics Comm. 194 (2001) 167.
\bibitem{Armenia} AVETISSIAN H.K. and MKRTCHIAN G.F., Phys. Rev.E 65 (2002) 046505.
\bibitem{MBE} ARECCHI F.T., BONIFACIO R., IEEE Quantum Electron. 1 (1965) 169.
\bibitem{BP} BONIFACIO R. and PREPARATA G., Phys. Rev.A 2 (1970) 336.
\end{thebibliography}
\end{document}